\documentstyle[prl,aps,12pt]{revtex}
\begin{document}
\draft
\title{\noindent{\large {\tt {} \hfill IFUSP/P-1118}}\\
\vskip 1.0cm Non-linear electromagnetic interactions in thermal QED}
\author{F. T. Brandt and J. Frenkel}
\address{Instituto de F\'\i sica, Universidade de S\~ao Paulo,\\
S\~ao Paulo, 01498 SP, Brasil}

\author{}

\maketitle

\begin{abstract}
We examine the behavior of the non-linear interactions between
electromagnetic fields at high temperature. It is shown that, in general,
the $\log (T)$ dependence on the temperature of the Green functions is
simply related to their UV behavior at zero-temperature. We argue that the
effective action describing the nonlinear thermal electromagnetic
interactions has a finite limit as $T\rightarrow \infty $. This thermal
action approaches, in the long wavelength limit, the negative of the
corresponding zero-temperature action.
\end{abstract}


The effective thermal action due to the electron-positron box, which is
fourth order in the electromagnetic field, has been studied previously in
the literature \cite{Dittrich,LoeweRojas,TARRACH}. In reference \cite
{BrandtFrenkelTaylor2} it was shown that this effective action has a finite
limit at high temperatures, when $T\rightarrow \infty $. The main purpose of
this paper is to extend the analysis in \cite{BrandtFrenkelTaylor2} on the
non-linear interactions, to all orders in the electromagnetic field.

At high temperatures, individual contributions contain power dependence on $T
$, but the $T^3$ and $T$ terms cancel by symmetry. Gauge invariance imposes
strong constraints, which lead to the cancellation of the $T^2$
contributions in QED \cite{FrenkelTaylor,BraatenPisarski}, except in the
case of the photon self-energy. Note that, in general, the dependence upon $T
$ for high $T$ is not necessarily connected to the UV divergence or
convergence of the zero-temperature amplitude. In QCD, for example, all the $%
N$-gluon functions behave like $T^2$, although they are UV finite for $N>4$.

This work addresses the problem of possible $\log (T)$ contributions. We
will present a simple argument showing that in thermal field theories, these
are related to the UV behavior of the Green functions at $T=0$.
Consequently, in the $N\geq 4$ photon Green functions, which are UV finite,
the $\log (T)$ contributions must be absent. We have verified this behavior
by explicit computation of the electron-positron 6-point function, in the
long wavelength limit of the external photons. Therefore, the effective
action describing the nonlinear interactions between electromagnetic fields
at high temperature must have a finite limit as $T\rightarrow \infty $. Like
the zero-temperature action, it must be gauge invariant. But this thermal
action may be more complicated, because it is not necessarily
Lorentz-invariant.

\begin{figure}
\setlength{\unitlength}{0.0079in}%
\begingroup\makeatletter\ifx\SetFigFont\undefined
\def\x#1#2#3#4#5#6#7\relax{\def\x{#1#2#3#4#5#6}}%
\expandafter\x\fmtname xxxxxx\relax \def\y{splain}%
\ifx\x\y   
\gdef\SetFigFont#1#2#3{%
  \ifnum #1<17\tiny\else \ifnum #1<20\small\else
  \ifnum #1<24\normalsize\else \ifnum #1<29\large\else
  \ifnum #1<34\Large\else \ifnum #1<41\LARGE\else
     \huge\fi\fi\fi\fi\fi\fi
  \csname #3\endcsname}%
\else
\gdef\SetFigFont#1#2#3{\begingroup
  \count@#1\relax \ifnum 25<\count@\count@25\fi
  \def\x{\endgroup\@setsize\SetFigFont{#2pt}}%
  \expandafter\x
    \csname \romannumeral\the\count@ pt\expandafter\endcsname
    \csname @\romannumeral\the\count@ pt\endcsname
  \csname #3\endcsname}%
\fi
\fi\endgroup
\begin{picture}(546,210)(-30,300)
%
%
\thicklines
\put(160,400){\circle{70}}
\put(490,400){\circle{70}}
\put(155,365){\vector( 1, 0){ 10}}
\put(195,395){\vector( 0, 1){ 10}}
\put(165,435){\vector(-1, 0){ 10}}
\put(125,405){\vector( 0,-1){ 10}}

\multiput(
95,465)(3.2,-3.2){13}{\makebox(0.4444,0.6667){\SetFigFont{10}{12}{rm}.}}
\multiput(225,465)(-3.2,-3.2){13}{\makebox(0.4444,0.6667){\SetFigFont{10}{12}{rm}.}}
\multiput(225,335)(-3.2,3.2){13}{\makebox(0.4444,0.6667){\SetFigFont{10}{12}{rm}.}}
\multiput(
95,335)(3.2,3.2){13}{\makebox(0.4444,0.6667){\SetFigFont{10}{12}{rm}.}}
\multiput(425,465)(3.2,-3.2){13}{\makebox(0.4444,0.6667){\SetFigFont{10}{12}{rm}.}}
\multiput(425,335)(3.2,3.2){13}{\makebox(0.4444,0.6667){\SetFigFont{10}{12}{rm}.}}
\multiput(555,465)(-3.2,-3.2){13}{\makebox(0.4444,0.6667){\SetFigFont{10}{12}{rm}.}}
\multiput(555,335)(-3.2,3.2){13}{\makebox(0.4444,0.6667){\SetFigFont{10}{12}{rm}.}}
\put(495,365){\vector(-1, 0){ 10}}
\put(455,395){\vector( 0, 1){ 10}}
\put(485,435){\vector( 1, 0){ 10}}
\put(525,405){\vector( 0,-1){ 10}}
%
\put( 60,315){\makebox(0,0)[lb]{\smash{\SetFigFont{12}{24.0}{rm}{\it k}}}}
\put( 70,311){\makebox(0,0)[lb]{\smash{\SetFigFont{8}{14.4}{rm}1}}}
\put( 80,315){\makebox(0,0)[lb]{\smash{\SetFigFont{12}{24.0}{rm},}}}
\put(95,315){\makebox(0,0)[lb]{\smash{\SetFigFont{12}{24.0}{rm}$\mu$}}}
\put(105,311){\makebox(0,0)[lb]{\smash{\SetFigFont{8}{14.4}{rm}1}}}
\put( 60,480){\makebox(0,0)[lb]{\smash{\SetFigFont{12}{24.0}{rm}{\it k}}}}
\put( 70,476){\makebox(0,0)[lb]{\smash{\SetFigFont{8}{14.4}{rm}2}}}
\put( 80,480){\makebox(0,0)[lb]{\smash{\SetFigFont{12}{24.0}{rm},}}}
\put(95,480){\makebox(0,0)[lb]{\smash{\SetFigFont{12}{24.0}{rm}$\mu$}}}
\put(105,476){\makebox(0,0)[lb]{\smash{\SetFigFont{8}{14.4}{rm}2}}}
\put(220,480){\makebox(0,0)[lb]{\smash{\SetFigFont{12}{24.0}{rm}{\it k}}}}
\put(230,476){\makebox(0,0)[lb]{\smash{\SetFigFont{8}{14.4}{rm}3}}}
\put(240,480){\makebox(0,0)[lb]{\smash{\SetFigFont{12}{24.0}{rm},}}}
\put(255,480){\makebox(0,0)[lb]{\smash{\SetFigFont{12}{24.0}{rm}$\mu$}}}
\put(265,476){\makebox(0,0)[lb]{\smash{\SetFigFont{8}{14.4}{rm}3}}}
\put(220,315){\makebox(0,0)[lb]{\smash{\SetFigFont{12}{24.0}{rm}{\it k}}}}
\put(230,311){\makebox(0,0)[lb]{\smash{\SetFigFont{8}{14.4}{rm}{\it N}}}}
\put(241,315){\makebox(0,0)[lb]{\smash{\SetFigFont{12}{24.0}{rm},}}}
\put(255,315){\makebox(0,0)[lb]{\smash{\SetFigFont{12}{24.0}{rm}$\mu$}}}
\put(265,311){\makebox(0,0)[lb]{\smash{\SetFigFont{8}{14.4}{rm}{\it N}}}}
\put(150,340){\makebox(0,0)[lb]{\smash{\SetFigFont{12}{24.0}{rm}{\it Q}}}}
\put(250,411){\makebox(0,0)[lb]{\smash{\SetFigFont{12}{34.8}{rm}.}}}
\put(251,400){\makebox(0,0)[lb]{\smash{\SetFigFont{12}{34.8}{rm}.}}}
\put(250,390){\makebox(0,0)[lb]{\smash{\SetFigFont{12}{34.8}{rm}.}}}
\put(390,315){\makebox(0,0)[lb]{\smash{\SetFigFont{12}{24.0}{rm}{\it k}}}}
\put(400,311){\makebox(0,0)[lb]{\smash{\SetFigFont{8}{14.4}{rm}1}}}
\put(410,315){\makebox(0,0)[lb]{\smash{\SetFigFont{12}{24.0}{rm},}}}
\put(425,315){\makebox(0,0)[lb]{\smash{\SetFigFont{12}{24.0}{rm}$\mu$}}}
\put(435,311){\makebox(0,0)[lb]{\smash{\SetFigFont{8}{14.4}{rm}1}}}
\put(390,480){\makebox(0,0)[lb]{\smash{\SetFigFont{12}{24.0}{rm}{\it k}}}}
\put(400,476){\makebox(0,0)[lb]{\smash{\SetFigFont{8}{14.4}{rm}2}}}
\put(410,480){\makebox(0,0)[lb]{\smash{\SetFigFont{12}{24.0}{rm},}}}
\put(425,480){\makebox(0,0)[lb]{\smash{\SetFigFont{12}{24.0}{rm}$\mu$}}}
\put(435,476){\makebox(0,0)[lb]{\smash{\SetFigFont{8}{14.4}{rm}2}}}
\put(551,480){\makebox(0,0)[lb]{\smash{\SetFigFont{12}{24.0}{rm}{\it k}}}}
\put(561,476){\makebox(0,0)[lb]{\smash{\SetFigFont{8}{14.4}{rm}3}}}
\put(571,480){\makebox(0,0)[lb]{\smash{\SetFigFont{12}{24.0}{rm},}}}
\put(586,480){\makebox(0,0)[lb]{\smash{\SetFigFont{12}{24.0}{rm}$\mu$}}}
\put(596,476){\makebox(0,0)[lb]{\smash{\SetFigFont{8}{14.4}{rm}3}}}
\put(551,315){\makebox(0,0)[lb]{\smash{\SetFigFont{12}{24.0}{rm}{\it k}}}}
\put(561,311){\makebox(0,0)[lb]{\smash{\SetFigFont{8}{14.4}{rm}{\it N}}}}
\put(572,315){\makebox(0,0)[lb]{\smash{\SetFigFont{12}{24.0}{rm},}}}
\put(586,315){\makebox(0,0)[lb]{\smash{\SetFigFont{12}{24.0}{rm}$\mu$}}}
\put(596,311){\makebox(0,0)[lb]{\smash{\SetFigFont{8}{14.4}{rm}{\it N}}}}
\put(580,410){\makebox(0,0)[lb]{\smash{\SetFigFont{12}{34.8}{rm}.}}}
\put(581,400){\makebox(0,0)[lb]{\smash{\SetFigFont{12}{34.8}{rm}.}}}
\put(580,390){\makebox(0,0)[lb]{\smash{\SetFigFont{12}{34.8}{rm}.}}}
\put(480,340){\makebox(0,0)[lb]{\smash{\SetFigFont{12}{24.0}{rm}{\it Q}}}}
\end{picture}
\caption[f1]{\label{f1}{Diagrams which contribute to the nonlinear
interactions between electromagnetic fields.
Dotted lines represent photons, and
solid lines stand for electrons or positrons.}}
\end{figure}
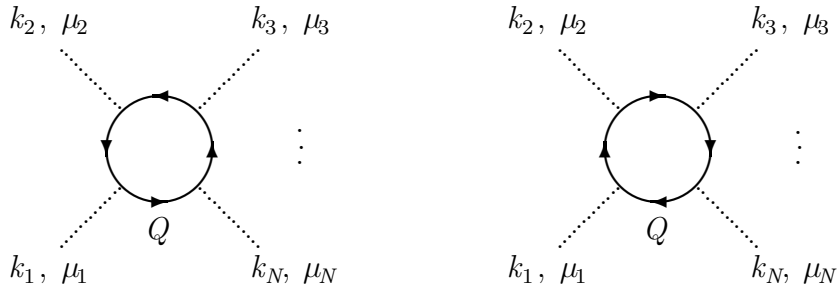

Typical graphs contributing to the nonlinear electromagnetic interactions
are shown in Fig. \ref{f1}. We must consider only diagrams with an even
number of external photon lines, since for $N$-odd, their contribution
vanishes by charge conjugation.

In order to discuss the logarithmic temperature dependence, we use the
analytically continued imaginary-time thermal perturbation theory\cite
{Kapusta}. Then, we can express the complete thermal amplitude, which
includes the zero-temperature part, in the form
\begin{equation}
\label{eq1}A^{\mu _1\mu _2\cdots \mu _N}({\bf k}_i,k_i^0,T)=M^\epsilon \
T\,\sum_{Q_0=\pi iT(2\,n+1)}\int {\rm d}^{3-\epsilon }{\bf Q}F^{\mu _1\mu
_2\cdots \mu _N}(Q^0,{\bf Q},{\bf k}_i,k_i^0).
\end{equation}
Here $M$ is the UV renormalization scale, $k_i^0/2\pi i\,T$ are integers and
$n$ runs over all integers. For fixed n, the {\bf Q}-integral is UV finite,
having no poles at $\epsilon =0$.

Our argument requires the identification of those terms which can yield a
pole at $\epsilon =0$, when performing in (\ref{eq1}) the summation over the
frequencies $Q^0=i\pi T(2n+1)$. To this end, we consider the high
temperature limit
\begin{equation}
\label{eq2}T>>|{\bf k}_i|,m
\end{equation}
($m$ is the electron mass) and examine a relevant contribution involving a
sum like\cite{BrandtFrenkelTaylor1}
\begin{equation}
\label{eq3}S=T\sum_{n=-\infty }^\infty \frac 1{|Q^0+k^0|^{1+\epsilon }}.
\end{equation}
Here $k^0$ is some linear combination of the external energies, with
integral coefficients. We can thus set: $k^0=2\pi ilT$, where $l$ is some
integer. Considering the contributions from the regions $|n|<|l|$ and $%
|n|\geq |l|$, we obtain in the limit $\epsilon \rightarrow 0$ the expression
\begin{equation}
\label{eq4}
\begin{array}{llll}
S & = &
{\displaystyle {T^{-\epsilon } \over \left( 2\pi \right) ^{1+\epsilon }}}
& \Biggl[\;2\Psi \left( |l+1/2|\right) -2\Psi \left( |l+1/2|-|l|\right) +
{\displaystyle {1 \over |l+1/2|}}+\Biggr.\\
&  &  & \;\;\;\zeta \left(1+\epsilon ,l+1/2\right)
                 +\zeta \left(1+\epsilon ,-l-1/2\right)\Biggl.\;\Biggr]
\end{array}
\end{equation}
where $\Psi $ is the Euler psi-function and $\zeta (\alpha ,z)$ denotes the
Riemann zeta-function \cite{Gradshteyn}
\begin{equation}
\label{eq5}\zeta (\alpha ,z)=\sum_{n=0}^\infty \frac 1{(n+z)^\alpha }.
\end{equation}
The only singular point of this function occurs as $\alpha \rightarrow 1$,
where it obeys the relation
\begin{equation}
\label{eq6}\zeta (1+\epsilon ,z)=\frac 1\epsilon -\Psi (z).
\end{equation}
Hence, using Eq. (\ref{eq4}), we see that the divergent part of $S$ is given
simply by
\begin{equation}
\label{eq7}S_\epsilon =\frac 1{\pi \epsilon }.
\end{equation}
This contribution arises from the summation over the region $|n|>>|l|$, i.e.
where $|Q_0|>>|k_0|$. It is associated with the leading term obtained by
expanding (\ref{eq3}) in powers of $k^0/Q^0$.

Consequently, for the purpose of evaluating the pole part of the complete
amplitude, we {\it can} expand the integral in (\ref{eq1}) in powers of $%
Q_0^{-1}$. Proceeding systematically in this way, and identifying all terms
proportional to $(Q^0)^{-1-\epsilon }$, we obtain contributions of the form
\begin{equation}
\label{eq8}A^{\mu _1\mu _2\cdots \mu _N}({\bf k}_i,k_i^0,T)=\sum_{n=-\infty
}^\infty \left[ \left( \frac MT\right) ^\epsilon \ f^{\mu _1\mu _2\cdots \mu
_N}({\bf k}_i,k_i^0,\epsilon )\ \frac 1{|2\,n+1|^{1+\epsilon }}+{\cal O}%
(T^{-1-\epsilon }n^{-2-\epsilon })\right] .
\end{equation}
If $f^{\mu _1\mu _2\cdots \mu _N}({\bf k}_i,k_i^0,0)$ is nonzero, this sum
diverges and the corresponding zero-temperature Euclidean field theory would
be UV divergent, having a pole like $1/\epsilon $. In this case we obtain
\begin{equation}
\label{eq9}A_\epsilon ^{\mu _1\mu _2\cdots \mu _N}({k}_i,T)=f^{\mu _1\mu
_2\cdots \mu _N}({k}_i,0)\left[ \frac 1\epsilon +
\log\left(\frac MT\right)\right] .
\end{equation}
We note that the $\log (T)$ term always combines with the $\log (M)$ term to
yield a $\log (M/T)$ contribution. Something similar happens for the photon
self-energy in QED, or the gluon two-point function in QCD \cite
{BrandtFrenkelTaylor1,Weldon}. Since $f^{\mu _1\mu _2\cdots \mu _N}({k}_i,0)$
is gauge and Lorentz invariant, we expect the $\log (T)$ contributions to be
Lorentz invariant quantities, despite the presence of the heat bath. This
was argued explicitly in Ref. \cite{BrandtFrenkelTaylor2}.

Because the Green functions with $N\geq 4$ external photon lines are UV
convergent, the functions $f^{\mu _1\mu _2\cdots \mu _N}({k}_i,0)$ must
vanish. Consequently, the $\log (T)$ terms should be absent in the
non-linear electromagnetic interactions at high temperature.

We now verify by explicit computation, the cancellation of the $\log (T) $
contributions to the 6-point photon function, which is UV finite. For
simplicity we consider here the long wavelength limit of the external
photons, when ${\bf k_i}\rightarrow 0$.

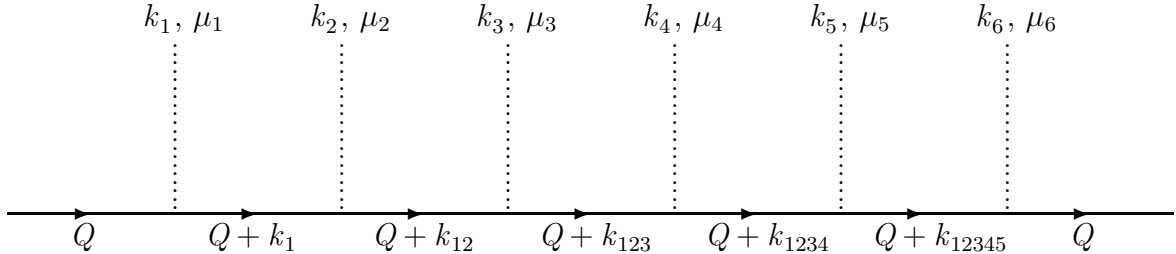
\begin{figure}[t]
\setlength{\unitlength}{0.0109in}%
\begingroup\makeatletter\ifx\SetFigFont\undefined
\def\x#1#2#3#4#5#6#7\relax{\def\x{#1#2#3#4#5#6}}%
\expandafter\x\fmtname xxxxxx\relax \def\y{splain}%
\ifx\x\y   
\gdef\SetFigFont#1#2#3{%
  \ifnum #1<17\tiny\else \ifnum #1<20\small\else
  \ifnum #1<24\normalsize\else \ifnum #1<29\large\else
  \ifnum #1<34\Large\else \ifnum #1<41\LARGE\else
     \huge\fi\fi\fi\fi\fi\fi
  \csname #3\endcsname}%
\else
\gdef\SetFigFont#1#2#3{\begingroup
  \count@#1\relax \ifnum 25<\count@\count@25\fi
  \def\x{\endgroup\@setsize\SetFigFont{#2pt}}%
  \expandafter\x
    \csname \romannumeral\the\count@ pt\expandafter\endcsname
    \csname @\romannumeral\the\count@ pt\endcsname
  \csname #3\endcsname}%
\fi
\fi\endgroup
\begin{picture}(560,135)(120,415)
%
%
\thicklines
\put(120,440){\vector( 1, 0){ 40}}
\put(160,440){\vector( 1, 0){ 80}}
\multiput(200,520)(0.0,-4.1){20}{\makebox(0.9,1.3){\SetFigFont{10}{12}{rm}.}}
\put(185,530){\makebox(0,0)[lb]{\smash{\SetFigFont{12}{14.4}{it}k}}}
\put(193,527){\makebox(0,0)[lb]{\smash{\SetFigFont{8}{8.0}{rm}1}}}
\put(200,530){\makebox(0,0)[lb]{\smash{\SetFigFont{12}{14.4}{rm}, $\mu_1$}}}
\put(150,425){\makebox(0,0)[lb]{\smash{\SetFigFont{12}{14.4}{it}Q}}}
\put(240,440){\vector( 1, 0){ 80}}
\multiput(280,520)(0.0,-4.1){20}{\makebox(0.9,1.3){\SetFigFont{10}{12}{rm}.}}
\put(265,530){\makebox(0,0)[lb]{\smash{\SetFigFont{12}{14.4}{it}k}}}
\put(273,527){\makebox(0,0)[lb]{\smash{\SetFigFont{8}{8.0}{rm}2}}}
\put(280,530){\makebox(0,0)[lb]{\smash{\SetFigFont{12}{14.4}{rm}, $\mu_2$}}}
\put(230,425){\makebox(0,0)[lb]{\smash{\SetFigFont{12}{14.4}{rm}+}}}
\put(215,425){\makebox(0,0)[lb]{\smash{\SetFigFont{12}{14.4}{it}Q}}}
\put(245,425){\makebox(0,0)[lb]{\smash{\SetFigFont{12}{14.4}{it}k}}}
\put(253,422){\makebox(0,0)[lb]{\smash{\SetFigFont{8}{8.0}{rm}1}}}
\put(320,440){\vector( 1, 0){ 80}}
\multiput(360,520)(0.0,-4.1){20}{\makebox(0.9,1.3){\SetFigFont{10}{12}{rm}.}}
\put(345,530){\makebox(0,0)[lb]{\smash{\SetFigFont{12}{14.4}{it}k}}}
\put(353,527){\makebox(0,0)[lb]{\smash{\SetFigFont{8}{8.0}{rm}3}}}
\put(360,530){\makebox(0,0)[lb]{\smash{\SetFigFont{12}{14.4}{rm}, $\mu_3$}}}
\put(310,425){\makebox(0,0)[lb]{\smash{\SetFigFont{12}{14.4}{rm}+}}}
\put(295,425){\makebox(0,0)[lb]{\smash{\SetFigFont{12}{14.4}{it}Q}}}
\put(325,425){\makebox(0,0)[lb]{\smash{\SetFigFont{12}{14.4}{it}k}}}
\put(333,422){\makebox(0,0)[lb]{\smash{\SetFigFont{8}{8.0}{rm}12}}}
\put(400,440){\vector( 1, 0){ 80}}
\multiput(440,520)(0.0,-4.1){20}{\makebox(0.9,1.3){\SetFigFont{10}{12}{rm}.}}
\put(425,530){\makebox(0,0)[lb]{\smash{\SetFigFont{12}{14.4}{it}k}}}
\put(433,527){\makebox(0,0)[lb]{\smash{\SetFigFont{8}{8.0}{rm}4}}}
\put(440,530){\makebox(0,0)[lb]{\smash{\SetFigFont{12}{14.4}{rm}, $\mu_4$}}}
\put(390,425){\makebox(0,0)[lb]{\smash{\SetFigFont{12}{14.4}{rm}+}}}
\put(375,425){\makebox(0,0)[lb]{\smash{\SetFigFont{12}{14.4}{it}Q}}}
\put(405,425){\makebox(0,0)[lb]{\smash{\SetFigFont{12}{14.4}{it}k}}}
\put(413,422){\makebox(0,0)[lb]{\smash{\SetFigFont{8}{8.0}{rm}123}}}
\put(480,440){\vector( 1, 0){ 80}}
\multiput(520,520)(0.0,-4.1){20}{\makebox(0.9,1.3){\SetFigFont{10}{12}{rm}.}}
\put(505,530){\makebox(0,0)[lb]{\smash{\SetFigFont{12}{14.4}{it}k}}}
\put(513,527){\makebox(0,0)[lb]{\smash{\SetFigFont{8}{8.0}{rm}5}}}
\put(520,530){\makebox(0,0)[lb]{\smash{\SetFigFont{12}{14.4}{rm}, $\mu_5$}}}
\put(470,425){\makebox(0,0)[lb]{\smash{\SetFigFont{12}{14.4}{rm}+}}}
\put(455,425){\makebox(0,0)[lb]{\smash{\SetFigFont{12}{14.4}{it}Q}}}
\put(485,425){\makebox(0,0)[lb]{\smash{\SetFigFont{12}{14.4}{it}k}}}
\put(493,422){\makebox(0,0)[lb]{\smash{\SetFigFont{8}{8.0}{rm}1234}}}
\put(560,440){\vector( 1, 0){ 80}}
\multiput(600,520)(0.0,-4.1){20}{\makebox(0.9,1.3){\SetFigFont{10}{12}{rm}.}}
\put(585,530){\makebox(0,0)[lb]{\smash{\SetFigFont{12}{14.4}{it}k}}}
\put(593,527){\makebox(0,0)[lb]{\smash{\SetFigFont{8}{8.0}{rm}6}}}
\put(600,530){\makebox(0,0)[lb]{\smash{\SetFigFont{12}{14.4}{rm}, $\mu_6$}}}
\put(550,425){\makebox(0,0)[lb]{\smash{\SetFigFont{12}{14.4}{rm}+}}}
\put(535,425){\makebox(0,0)[lb]{\smash{\SetFigFont{12}{14.4}{it}Q}}}
\put(565,425){\makebox(0,0)[lb]{\smash{\SetFigFont{12}{14.4}{it}k}}}
\put(573,422){\makebox(0,0)[lb]{\smash{\SetFigFont{8}{8.0}{rm}12345}}}
\put(640,440){\line( 1, 0){ 40}}
\put(630,425){\makebox(0,0)[lb]{\smash{\SetFigFont{12}{14.4}{it}Q}}}
\end{picture}

\nopagebreak
\caption[f2]{\label{f2}{A typical contribution to the
           forward scattering amplitude.
           Dotted lines represent photons, and
           solid lines stand for electrons or positrons.}}
\end{figure}

The analytically continued imaginary-time formalism can be formulated \cite
{Barton,FrenkelTaylor2} so as to express the thermal amplitude (having
subtracted the zero-temperature part) in terms of amplitudes for forward
scattering of a thermal electron or positron in an external electromagnetic
field, as depicted in Fig. \ref{f2}. There are $6!$ diagrams like this one,
which are obtained by permutations of the external momenta and
polarizations. The corresponding analytic expression has the form
\begin{equation}
\label{eq10}\frac{e^6}{\left( 2\pi \right) ^3}%
\displaystyle \int
_0^\infty \frac{q^2}{2q^0}\frac{{\rm d}q}{\exp \left( q^0/T\right) +1}\int
{\rm d}\Omega \sum_{\left( {\rm i\,j\,k\,l\,m\,n}\right) }B_{\left( {\rm %
i\,\,j\,k\,l\,m\,\,n}\right) }^{\mu _1\mu _2\mu _3\mu _4\mu _5\mu _6}\left(
k_1,k_2,k_3,k_4,k_5,k_6,Q\right) .
\end{equation}
Here $q=\left| {\bf Q}\right| $, $q^0=\left( q^2+m^2\right) ^{1/2}$, $\int
{\rm d}\Omega $ is an integral over the directions of ${\bf Q}$, and the sum
is over the permutations $\left( {\rm i\,j\,k\,l\,m\,n}\right) $ of $\left( 1%
{\rm \,2\,3\,4\,5\,6}\right) $. Each $B$ has a numerator which is a Dirac
trace containing projection operators $P\left( k\right) =\left[ \gamma \cdot
\left( Q+k\right) +m\right] $. For example
\begin{equation}
\label{eq11}B_{\left( 1{\rm \,2\,3\,4\,5\,6}\right) }^{\mu _1\mu _2\mu _3\mu
_4\mu _5\mu _6}=\frac{{\rm tr}\left[ P\left( 0\right) \ \gamma ^{\mu _1}\
P\left( k_1\right) \ \gamma ^{\mu _2}\ P\left( k_{12}\right) \ \gamma ^{\mu
_3}\ \cdots \ P\left( k_{12345}\right) \ \gamma ^{\mu _6}\right] }{\left(
2Q\cdot k_1+k_1^2\right) \left( 2Q\cdot k_{12}+k_{12}^2\right) \cdots \left(
2Q\cdot k_{12345}+k_{12345}^2\right) },
\end{equation}
where $k_{12}=k_1+k_2$, etc.

In order to obtain the high-temperature limit of Eq. (\ref{eq10}) we use the
expansion
\begin{equation}
\label{eq12}\left( 2Q\cdot k+k^2\right) ^{-1}=\left( 2Q\cdot k\right)
^{-1}\left[ 1-k^2\left( 2Q\cdot k\right) ^{-1}+k^4\left( 2Q\cdot k\right)
^{-2}+\cdots \right]
\end{equation}
in (\ref{eq11}). For our purposes, we can neglect the electron mass in (\ref
{eq10}). One thus gets in Eq. (\ref{eq11}) terms which are homogeneous in $Q$
of degrees 1, 0, -1, -2. The expansion cannot be taken further without
introducing spurious infra-red divergences. Since for each term in the sum
in Eq. (\ref{eq10}) there is a corresponding one with $Q\leftrightarrow -Q$,
terms which are odd in $Q$ will cancel. The terms of degree 0 would produce $%
{\cal O}(T^2)$. However, as we have already mentioned, these contributions
cancel as a consequence of gauge invariance. We have checked explicitly this
cancellation at the integrand level in Eq. (\ref{eq10}).

Terms of degree -2 would produce possible $\log (T)$ contributions. The
explicit computation of this logarithmic contribution is lengthy and can
only be performed using a computer program for symbolic manipulations. Even
with the help of a computer, one would find very difficult angular integrals
for general values of the external photon momenta. In the long wavelength
limit, one has to consider only the space components of Eq. (\ref{eq10}),
because any other component must vanish by gauge invariance. The angular
integrations can then be easily computed. There are in fact no angular
dependences left in Eq. (\ref{eq11}) apart from simple numerators like $%
Q^{i_1}\,Q^{i_2}\,\cdots \,Q^{i_n}$, where $n=0,2,4,6$ and $i_1$,$\cdots $, $%
i_n$ represent the space directions. Even in this relatively simple case the
number of terms resulting from the expansion of a single contribution such
as (\ref{eq11}) is 975. After adding the $6!$ permutations of all the above
contributions, and using the conservation of the photon momenta, we obtain
zero.

In conclusion, we discuss the finite contributions to the effective thermal
action at high temperature. To this end, we convert the frequency sum in the
complete amplitude (\ref{eq1}) to contour integrals, expressed at zero
chemical potential as \cite{Kapusta}

\begin{equation}
\label{eq13}A({\bf k}_i,k_i^0,T)=\frac 1{2\pi i}\int_{-i\infty +\delta
}^{i\infty +\delta }{\rm d}Q^0\left( I(Q^0,{\bf k}_i,k_i^0)+I(-Q^0,{\bf k}%
_i,k_i^0)\right) \left[ \frac 12-\frac 1{\exp \left( Q^0/T\right) +1}\right]
,
\end{equation}
where $I$ is given by the ${\bf Q}$-integral in Eq. (\ref{eq1}), and we have
omitted for simplicity of notation all Lorentz indices. The first term in
the square bracket of (\ref{eq13}) leads to the zero temperature Euclidean
amplitude, while the second one contains the temperature-dependent Fermi
distribution $N(Q^0/T)$.

We now evaluate the $Q^0$-integral by residues in terms of the poles inside
the contour. In general, these poles will be situated at $\left[ \left( {\bf %
Q}+{\bf k}\right) ^2+m^2\right] ^{1/2}+k^0$, where ${\bf k}$ and $k^0$
denote some linear combinations with integral coefficients, of the external
momenta and energies. An important simplification occurs in the long
wavelength limit ${\bf k}_i\rightarrow 0$, when the contributions from all
poles to the Fermi distribution factors become equal to
\begin{equation}
\label{new15}N\left( (q^0+k^0)/T\right) =N(q^0/T)=\frac 1{\exp {\left(
q^0/T\right) }+1},
\end{equation}
where $q^0=\left({\bf Q}^2+m^2\right)^{1/2}$.
Denoting by $R_{+}$ the residues at the poles of the function $F(Q^0,{\bf Q}%
,k_i^0)+F(-Q^0,{\bf Q,}k_i^0)$ (see Eq.
(\ref{eq1})), we get from (\ref{eq13}) the result
\begin{equation}
\label{new16}A(k_i^0,T)=\int {\rm d}^3{\bf Q}R_{+}\left( {\bf Q}%
,q^0,k_i^0\right) \left[ -\frac 12+\frac 1{\exp {\left( q^0/T\right) }+1}%
\right].
\end{equation}
Here we have set $\epsilon =0$, since the nonlinear Green functions are UV
finite. At this point, we can analytically continue the complete amplitude to
general values of the energies $k_i^0$.

We use next the important fact that the thermal contributions have a finite
limit as $T\rightarrow \infty $. Thus, we can take in (\ref{new16}) the
limit $N\left( q^0/T\right) \rightarrow 1/2$, without affecting the
convergence of the ${\bf Q}$-integral. It follows that in this case, the
nonlinear thermal action evaluated in the long wavelength limit, approaches
the negative of the zero-temperature action. In this limit, we expect
important effects due the collective behavior of the thermal medium.

\acknowledgements{We would like to thank ${\rm CNP_Q}$, Brasil, for a grant,
and Prof. J. C. Taylor for many valuable discussions}.

\end{document}